\documentclass[%
 reprint,
superscriptaddress,
showpacs,
 amsmath,amssymb,
prl,
]{revtex4-1}

\usepackage{graphicx}% Include figure files
\usepackage{dcolumn}% Align table columns on decimal point
\usepackage{bm}% bold math
\usepackage{color}
\usepackage{hyperref}
\hypersetup{backref,colorlinks=true,breaklinks,urlcolor=blue,linkcolor=blue,citecolor=blue}
\usepackage{siunitx}
\usepackage{array}
\usepackage{booktabs}
\usepackage[table]{xcolor}
\usepackage{ulem}

\begin{document}

\preprint{APS/123-QED}

\title{Observation of broad $d$-wave Feshbach resonances with a triplet structure}
\author{Yue Cui}
\author{Chuyang Shen}
\author{Min Deng}
\author{Shen Dong}
\author{Cheng Chen}
\affiliation{State Key Laboratory of Low Dimensional Quantum Physics, Department of Physics, Tsinghua University, Beijing 100084, China}
\author{Rong L\"u}
\affiliation{State Key Laboratory of Low Dimensional Quantum Physics, Department of Physics, Tsinghua University, Beijing 100084, China}
\affiliation{Collaborative Innovation Center of Quantum Matter, Beijing, China}
\author{Bo Gao}
\email{bo.gao@utoledo.edu}
\affiliation{Department of Physics and Astronomy, University of Toledo, Mailstop 111, Toledo, Ohio 43606, USA}
\author{Meng Khoon Tey}
\email{mengkhoon\_tey@mail.tsinghua.edu.cn}
\affiliation{State Key Laboratory of Low Dimensional Quantum Physics, Department of Physics, Tsinghua University, Beijing 100084, China}
\affiliation{Collaborative Innovation Center of Quantum Matter, Beijing, China}
\author{Li You}
\email{lyou@mail.tsinghua.edu.cn}
\affiliation{State Key Laboratory of Low Dimensional Quantum Physics, Department of Physics, Tsinghua University, Beijing 100084, China}
\affiliation{Collaborative Innovation Center of Quantum Matter, Beijing, China}


\date{\today}% It is always \today, today,

\begin{abstract}
High partial-wave ($l\ge2$) Feshbach resonance (FR) in an ultracold mixture of $^{85}$Rb-$^{87}$Rb atoms is investigated experimentally aided by a partial-wave insensitive analytic multichannel quantum-defect theory (MQDT). Two ``broad" resonances from coupling between $d$-waves in both the open and closed channels are observed and characterized. One of them shows a fully resolved triplet structure with splitting ratio well explained by the perturbation to the closed channel due to interatomic spin-spin interaction. These tunable ``broad" $d$-wave resonances, especially the one in the lowest-energy open channel, could find important applications in simulating $d$-wave coupling dominated many-body systems. In addition, we find that there is generally a time and temperature requirement, associated with tunneling through the angular momentum barrier, to establish and observe resonant coupling in nonzero partial waves.
\end{abstract}

\maketitle

Ultracold atoms with controllable interaction via Feshbach resonance (FR) have provided an ideal platform to study novel phenomena in few- and many-body physics~\cite{Chin2010FRreview}. While quantum gases often display smooth crossover behavior when crossing a $s$-wave FR~\cite{Leggett1980,Jochim2003BECofmolecule,Salomon2004BECBCS,Jin2004BECBCS,Ketterle2004FRcondensation}, they are predicted to exhibit complex order parameters and quantum phase transitions when driven across FRs of higher partial waves~\cite{Botelho2005,Andreev2005pwavetransition,Yip2005FermiSuperfluid,Gurarie2007pwavesuperfluid,Radzihovsky2009pwave}. High partial-wave FRs can enhance the nominally suppressed nonzero partial-wave interactions at low temperatures to the unitarity limit~\cite{Chin2010FRreview}. They give access to high partial-wave pairing which plays an important role in $p$-wave superfluidity in liquid $^{3}$He~\cite{Lee1997He3} or the proposed $d$-wave high-$T_c$ superconductors~\cite{Kirtley2000dwave}, and can significantly expand the platforms for cold atom based quantum simulations.

To coherently control high partial-wave interactions, it is crucial to find suitable high partial-wave FRs with small atom losses. Nearly lossless FRs have only been found in ``broad" \cite{Chin2010FRreview,Gao2011MQDTFR} $s$-wave resonances of fermionic alkali mixtures. All $p$-wave FRs observed to-date are accompanied by strong losses due to either two-body dipolar spin flip or three-body recombination, even for fermionic atomic species ~\cite{Jin2003pwaveK40,Salomon2004pwaveLi,Ketterle2005Li,Grimm2008LiK}. The prevailing wisdom has been to search for high partial-wave FRs in the lowest energy open channels to avoid exothermic dipolar loss, and to search for ``broad" high partial-wave FRs dominated by open channels to minimize the influence of the bound states. Unfortunately, the only alkali fermionic atoms, either $^6$Li or $^{40}$K or their mixtures, are void of $p$-wave FRs satisfying both criteria simultaneously~\cite{Gao2014LiKFR}.

A natural progression is from $p$-wave to $d$-wave resonances. ``Broad" $d$-wave FRs originating from the direct coupling between a $d$-wave open channel and a $d$-wave closed channel are expected to exhibit a triplet structure, akin to the doublet structure for $p$-wave FRs~\cite{Jin2004pwavedoublet}.  These $d$-wave FRs, however, have not been observed or identified before. Almost all FRs which are loosely referred as ``$d$-wave FRs'' previously arise from coupling between a $s$-wave open channel and a $d$-wave closed channel~\cite{Verhaar2002Rb87,Simoni2006KRb,Oberthaler2011Na,Tiemann2012LiNa,Cornish2013Rb85,Grimm2013Cs,Cornish2013Rb85Cs,Cornish2014Rb87Cs}. These are actually $s$-wave FRs (scattering) induced by a $d$-wave bound state. A distinguishing experimental feature is that atom losses caused by these FRs do not disappear even at zero temperature unlike real $p$-wave and $d$-wave FRs. One exception is the FR observed in $^{52}$Cr atoms originating from coupling between a $d$-wave open channel and a $s$-wave bound state~\cite{Pfau2005dwaveCr,Gorceix2009dwaveCr}. Both cases described above exhibit one resonance peak.

\begin{figure*}[ht]
    \includegraphics[width=0.9\linewidth]{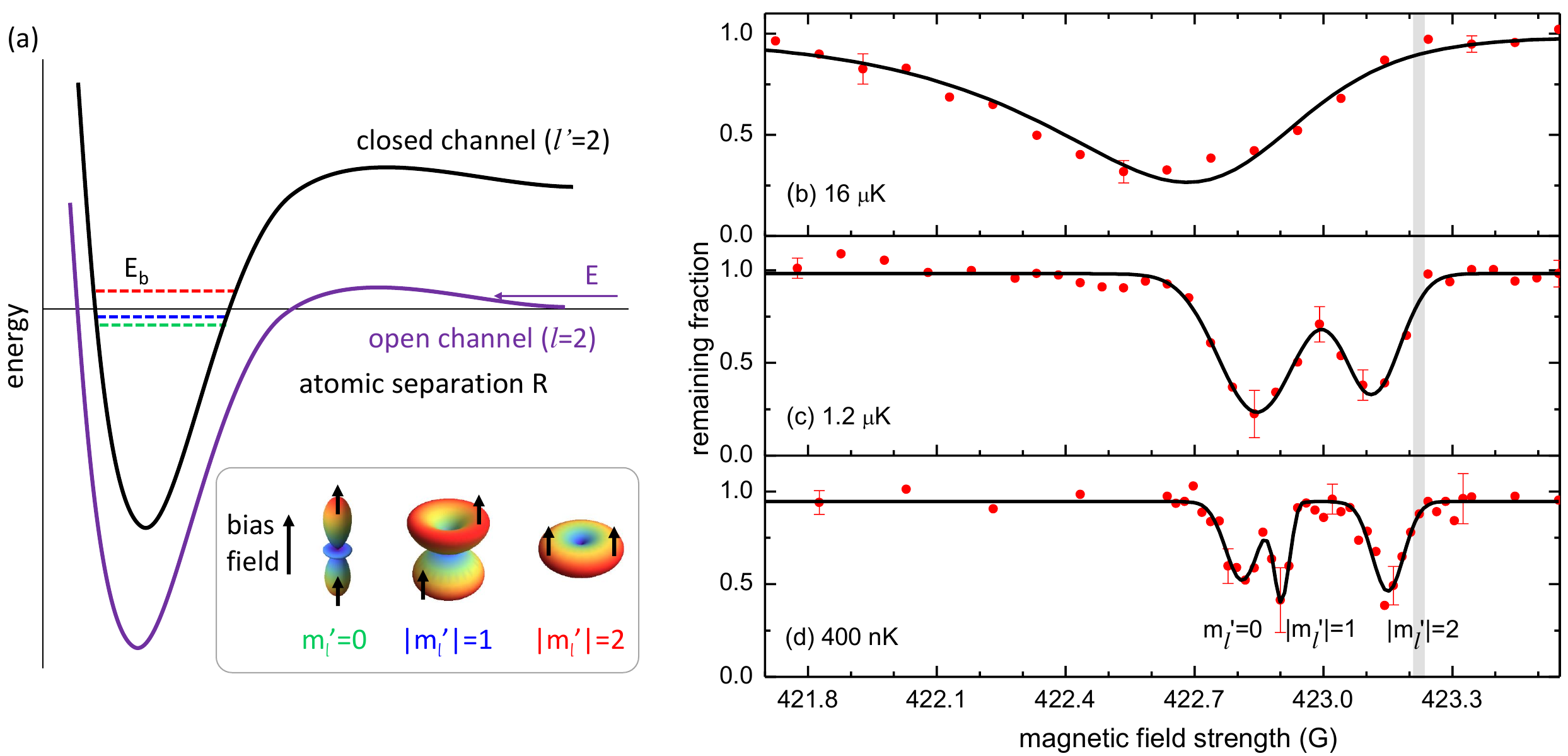}
    \caption{(Color online) The $d$-wave Feshbach resonance. (a) A simplified two-channel illustration for the origin of the triplet structure in $d$-wave FRs. Inset shows the angular part of the molecular wave function (spherical harmonic $|Y_2^{m_l'}|^2$), with arrows illustrating the spins of the valence electrons aligned with the bias magnetic field (for the spin-triplet closed channel). In this example, the $m_l'=0$ ($|m_l'|=2$) state has lower (higher) energy because the magnetic dipole-dipole interaction is dominated by the head-to-tail attractive (side-by-side repulsive) configuration. The splitting of the bound states gives rise to a triplet FR structure when the open channel is also $d$-wave. (b)-(d) show the remaining fraction of $^{85}$Rb (normalized to the baseline value in off-resonance regions) after coexisting with $^{87}$Rb atoms (b) for 1\,s at \SI{16}{\micro\kelvin}, (c) for 1\,s at 1.2\,$\mu$K, and (d) for 1.6\,s at 400\,nK. The number of $^{85}$Rb and $^{87}$Rb atoms in the off-resonance regions are, respectively, about $2\times10^{5}$ and $4\times10^{5}$ for (b), $1.2\times10^{4}$ and $1.2\times10^{5}$ for (c), and $3.5\times10^{3}$ and $5.5\times10^{4}$ for (d). The characteristic triplet structure for $l=l'=2$ is clearly visible in (d). The shaded gray region indicates the position at which the $|m_l'|=2$ bound states become degenerate with the threshold of the open channel. All data points are averages over five measurements, and error bars show the typical standard deviations. The eye-guiding black solid lines in (b) [(c),(d)] are fits to the data using asymmetric double Sigmoidal [multiple-peak] function.
    }
    \label{splitting}
\end{figure*}

This Letter reports the first observation of two ``broad" $d$-wave FRs, arising from the direct coupling between an open channel $d$-wave and a closed channel $d$-wave. We find a triplet structure with a peak-separation ratio in good agreement with interatomic spin-spin interaction~\cite{Tiemann2010spinspin} induced level splitting. One of the resonances is in the lowest-energy open channel and is thus free from two-body dipolar spin flip. This work is stimulated and guided by a partial-wave insensitive multichannel quantum-defect theory (MQDT) based on the analytic eigenfunctions for diatomic systems with long-range van der Waals (vdW) potential tail~\cite{Gao1998r6solutions,Gao2005MQDT,Gao2009singlechannel,Gao2011MQDTFR}. The MQDT predicts many ``broad'' $d$- and $f$-wave FRs in the mixture of $^{85}$Rb and $^{87}$Rb atoms. However, we find that to realize true $f$-wave coupling, which gives the $f$-wave quartet, requires a `tunneling time' (the time for the entrance wave function to build up inside the centrifugal barrier in order to feel the existence of the closed channel) longer than our present setup can provide at ultracold temperatures.

The experimental procedure for preparing an ultracold mixture of $^{85}$Rb and $^{87}$Rb atoms has been described previously~\cite{You2016Rb85Rb87}. In brief, the experiment starts with loading of both atomic isotopes into a magneto-optical trap (MOT). This is followed by optical pumping of the atoms into their corresponding low-field seeking states, $^{85}$Rb$|f=2,m_f=-2\rangle$ and $^{87}$Rb $|1, -1\rangle$, before transferring them into a magnetic quadrupole trap. After forced microwave evaporation of $^{87}$Rb atoms, which sympathetically cools $^{85}$Rb atoms, the mixture is loaded into a crossed optical dipole trap formed by two horizontal 1064-nm light beams with $1/e^{2}$ waists of $\sim$\SI{35}{\micro\meter} and $\sim$\SI{120}{\micro\meter}, and of powers of 2.7\,W and 3\,W, respectively.
At this stage, we have typically $8\times10^{6}$ $^{85}$Rb and $9\times10^{6}$ $^{87}$Rb atoms at a temperature of $\sim$25\,$\mu$K. The atoms are subsequently cooled to temperatures ranging from 400\,nK to \SI{16}{\micro\K} by reducing the powers of both trapping beams, followed by rf-adiabatic-passage transfers \cite{Bloch1997adiabaticpassage} to the desired spin states if required. Inter-isotope FRs are then detected by monitoring the fractional losses of both isotopes after ramping the magnetic field to a desired value and holding the mixture for a certain amount of time. The presence of a FR is manifested by enhanced atom losses.

The predictive power of the analytic MQDT for non-zero partial-wave FRs has been proven by our previous observations of ``broad'' $p$-wave FRs in the $^{85}$Rb and $^{87}$Rb mixture~\cite{You2016Rb85Rb87}. With the discovery of the $d$-wave FRs in this Letter, the territory governed by the analytic MQDT is further expanded. These $d$-wave FRs we observe differ from all previously reported $d$-wave-related FRs~\cite{Verhaar2002Rb87,Simoni2006KRb,Oberthaler2011Na,Tiemann2012LiNa,Cornish2013Rb85,Grimm2013Cs,Cornish2013Rb85Cs,Cornish2014Rb87Cs,Pfau2005dwaveCr,Gorceix2009dwaveCr} that arise from coupling between a $l=0$ ($l=2$) open channel and a $l'=2$ ($l'=0$) closed channel. The $|\Delta l|=2$ coupling behind those FRs is facilitated by the weak anisotropic interatomic spin-spin interaction~\cite{Tiemann2010spinspin} which represents the combined effect of the magnetic dipole-dipole %~\cite{Verhaar1988dipole,Axelsson1995dipole}
and the second order spin-orbit interaction between the valence electrons. %~\cite{Krauss1996spinorbit,Julienne2000spinorbit}.
For the ($\Delta l=0$) $d$-wave FR we report, the coupling arises instead from the much stronger isotropic electronic interaction~\cite{Chin2010FRreview}, which can potentially give ``broader'' $d$-wave FRs. The distinguishing signature of these two types of resonance lies at the fact that the latter shows a triplet structure instead of a singlet for the former.

Figure~\ref{splitting}(a) shows a simplified two-channel illustration for the origin of the triplet structure. The spin-spin interaction between the valence electrons perturbs the degenerate $l'=2$ molecular bound states and consequently splits them into three according to the azimuthal quantum number $|m_l'|=0,1,2$. For a $s$-wave ($l=0$) in the open channel, only one FR can be observed as it can only couple to one of the $m_l'$ bound states, because the total azimuthal quantum number $m_l$ + $M_{F}$ ($M_{F}$ = $m_{f,85}$+$m_{f,87}$) is conserved according to the coupling mechanisms involved. Likewise, a FR with a $d$-wave open channel and a $s$-wave closed channel also exhibits one peak. On the contrary, for FRs that arise from coupling through electronic (exchange) interactions, the selection rules are  $l=l'$, $m_l=m_l'$, $M_F=M_F'$. This, together with the fact that both the open and the closed channels can take $2l$+1 values of $m_l$, leads to a total of $l$+1 FR peaks (FRs with equal $|m_l|$ ($m_l\neq0$) are doubly degenerate). Hence, a triplet structure for the $d$-wave is expected.

Figures~\ref{splitting}(b)-(d) show the loss spectra around a $d$-wave FR near 423.0\,G in the $^{85}$Rb$|2, -2\rangle$+$^{87}$Rb$|1, -1\rangle$ open channel at different temperatures. The remaining fraction of $^{85}$Rb atoms are plotted as a function of the magnetic field after coexisting with $^{87}$Rb atoms for 1\,s at \SI{16}{\micro\kelvin} (b), for 1\,s at \SI{1.2}{\micro\kelvin} (c), and for 1.6\,s at 400\,nK (d). Overall, the loss features are observed to shift towards higher magnetic fields and a triplet structure emerges as the temperature of the atoms is lowered. The shift can be understood by noting first that, for this particular FR, the magnetic moment for the bound molecule is less negative than that of the two atoms in the open channel. The bound states thus move upward with respect to the threshold of the open channel with decreasing magnetic field. Second, at higher temperatures, atoms can access bound states that lie above the threshold of the open
channel supplemented by their kinetic energy. Together, these two factors shift the maximal loss towards lower magnetic fields at higher temperatures as observed.

To quantitatively understand the splitting of the $d$-wave resonance, we study the perturbation of the bound states due to the magnetic dipole-dipole interaction between the valence electrons of the alkali atoms~\cite{Jin2004pwavedoublet}. The perturbing interaction can be expressed in the tensor operator form as~\cite{Stoof1993dipole}
\begin{equation}\label{dipoletensor}
H'=-\frac{\alpha^2\sqrt{6}}{R^3}\sum_{q=-2}^2 (-1)^q C_{q}^{2}(s_{1}\otimes s_{2})_{-q}^{2},
\end{equation}
in atomic units, where $\alpha$ is the fine structure constant, $R$ the separation between the two atoms, $C_{q}^{2}$ the reduced spherical harmonic defined as $C_{q}^{k}(\theta,\phi)=\sqrt{4\pi/(2k+1)}Y_{kq}(\theta,\phi)$, and $(s_{1}\otimes s_{2})_{-q}^{2}$ the second-rank tensor formed from the spin operators of atom 1 and 2. The matrix elements $H'_{ij}$ are given by
\begin{eqnarray}\label{matrixelement}
-\alpha^2\sqrt{6}\sum_{q=-2}^2 (-1)^q \langle{l' m^\prime_{li}}|C_{q}^{2}|l' m^\prime_{lj}\rangle
\langle\phi_{b}|\frac{(s_{1}\otimes s_{2})_{-q}^{2}}{R^3}|\phi_{b}\rangle,\hskip 15pt
\end{eqnarray}
where $|\phi_{b}\rangle$ is the coupled-channel wave function for the bound state with coupling to the continuum. It takes the same form for different $m_l'$ within the first order approximation. For $d$-wave FRs with open and closed channels of the same $M_{F}$, only the $q = 0$ terms contribute. Thus $H'$ is diagonal and the energy shifts of the bound states are
\begin{eqnarray}\label{energyshift}
\Delta E_{m^\prime_{l}}=-\alpha^2\sqrt{6}\langle{2,m_l^{\prime}}|C_{0}^{2}|2,m_l^{\prime}\rangle
\langle\phi_{b}|\frac{(s_{1}\otimes s_{2})_{0}^{2}}{R^3}|\phi_{b}\rangle,\hskip 12pt
\end{eqnarray}
differing only in the multiplying factor $\langle{2,m_l^{\prime}}|C_{0}^{2}|2,m_l^{\prime}\rangle$. Since $\langle{2,0}|C_{0}^{2}|2,0\rangle=\frac{2}{7}$, $\langle{2,\pm1}|C_{0}^{2}|2,\pm1\rangle=\frac{1}{7}$, $\langle{2,\pm2}|C_{0}^{2}|2,\pm2\rangle=-\frac{2}{7}$,
the triplet structure is expected to exhibit a splitting ratio of $(\Delta E_{m_{l}'=0}-\Delta E_{|m_{l}'|=1})/(\Delta E_{|m_{l}'|=1}-\Delta E_{|m_{l}'|=2})=1/3$. The second-order spin-orbit interaction takes the same form as the magnetic dipole-dipole interaction in the spin dependant part~\cite{Tiemann2010spinspin}, and thus does not change the splitting ratio.

Experimentally, we measure an offset of 92(7)\,mG between the $|m_{l}'|= 0$ and 1 peaks, and an offset of 247(6)\,mG between the $|m_{l}'| = 1$ and 2, giving a ratio of about 1/2.7(3) which agrees well with the theoretical expectation.
The measured splitting ratio thus supports the origin of the triplet structure as due to spin-spin interaction, analogous to the widely studied p-wave doublet. Nevertheless, it represents an independent affirmation from latter since this ratio does not require the accurate knowledge of the bound state wave function or the magnitude of the second-order spin-orbit interaction.

\begin{figure}[b]
    \includegraphics[width=1\linewidth]{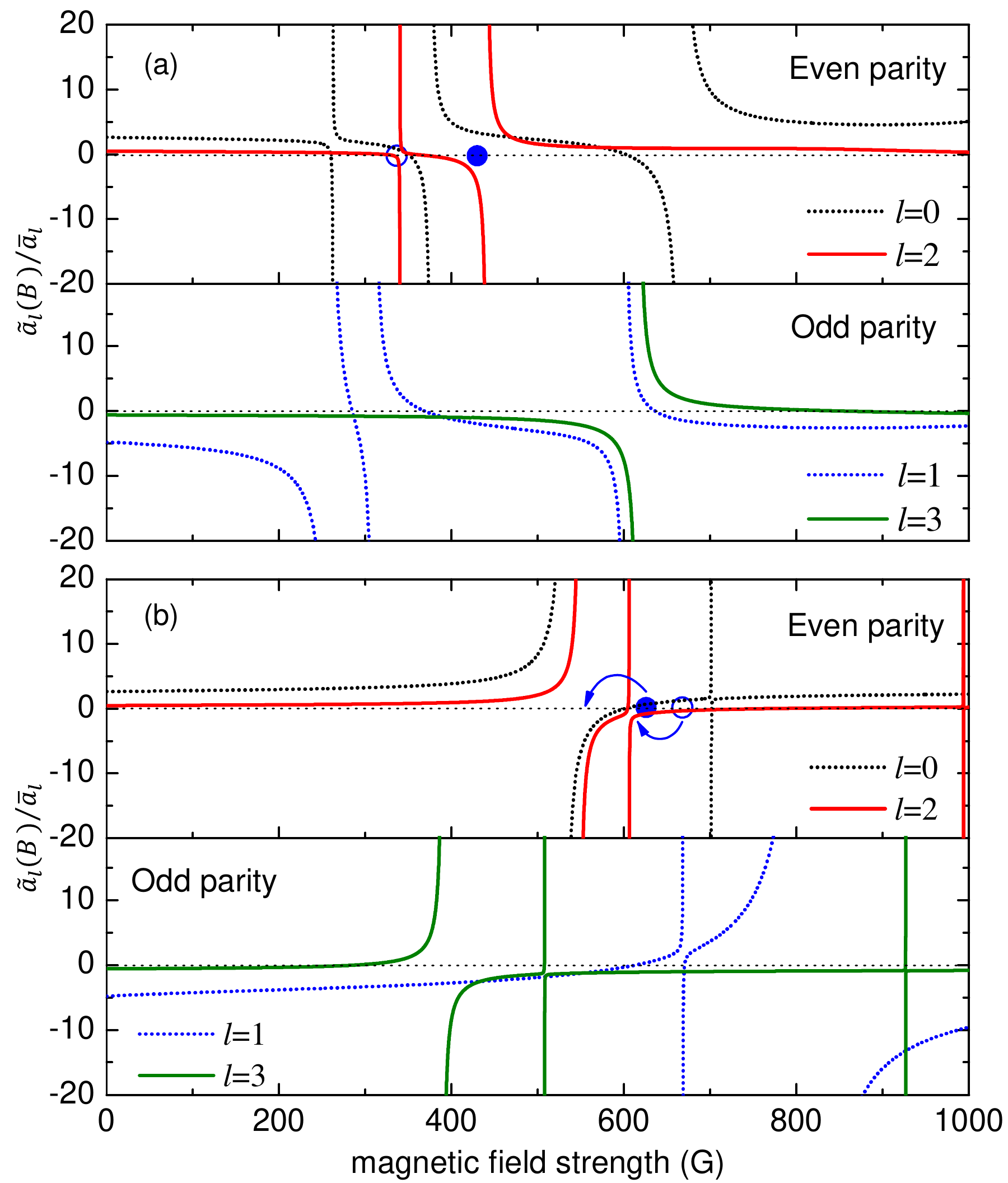}
    \caption{(Color online)
    Reduced generalized scattering lengths ${\tilde{a}_{l}(B)}/{\bar{a}_{l}}$ in the (a) $^{85}$Rb$|2, -2\rangle$+$^{87}$Rb$|1, -1\rangle$ and (b) $^{85}$Rb$|2, +2\rangle$+$^{87}$Rb$|1, +1\rangle$ scattering channels, for partial waves $l$ = 0 through $l$ = 3. The blue filled (empty) circles denote experimentally observed positions for the ``broad'' (``narrow") d-wave FRs. The arrows in (b) link the observed FRs to their corresponding predicted positions. The identification is done through coupled-channel calculations.}
    \label{rgsl}
\end{figure}

The $d$-wave FR we discuss above, together with a number of other $d$- and $f$-wave FRs, is predicted using a MQDT assisted with analytic solutions to long-range vdW potential in any partial waves~\cite{Gao1998r6solutions,Gao2005MQDT,Gao2009singlechannel,Gao2011MQDTFR}. Within the MQDT framework we adopt, the properties of FRs in all partial waves are determined from three parameters~\cite{Gao2005MQDT,Julienne2009threeparameter}: the vdW coefficient $C_6$, the singlet $s$-wave scattering length $a_{l = 0}^{S}$, and the triplet $s$-wave scattering length $a_{l = 0}^{T}$, besides the inherent atomic parameters such as hyperfine splitting. The FRs in the $^{85}$Rb-$^{87}$Rb mixture are calculated here using $C_{6}$ = 4710 a.u., $a_{l = 0}^{S}$ = 11.37 a.u., and $a_{l = 0}^{T}$ = 184.0 a.u., taken from Ref.~\cite{You2016Rb85Rb87}.

The MQDT provides for a unified parametrization of FRs in all partial waves. One of the important parameters is the derived resonance strength $\zeta_\textrm{res}$~\cite{Gao2011MQDTFR}, which is used to distinguish ``broad" resonances ($|\zeta_\textrm{res}|\gg1$) from the narrow ones ($|\zeta_\textrm{res}|\ll1$). The effective atomic interaction for the former ones follows the single-channel universal behaviour dominated by the open channel. Another parameter is the generalized scattering length ${\tilde{a}_{l}}$~\cite{Gao2009singlechannel,Gao2011MQDTFR} which is of dimension L$^{2l+1}$ for the $l$-th partial wave. For $l$ = 0 ($l$ =1), it coincides with the $s$-wave scattering length ($p$-wave scattering volume) of the effective-range approximation. When normalized by a mean scattering length $\bar{a}_{l}=\bar{a}_{sl}\beta_6^{2l+1}$~\cite{Gao2009singlechannel} (with $\beta_6=(2\mu C_6/\hbar^2)^{1/4}$, $\mu$ the reduced mass, and $\bar{a}_{sl}$ a $l$-dependent constant~\cite{aslformula}), ${\tilde{a}_{l}}$ for different partial waves can be conveniently plotted together.

Figures~\ref{rgsl}(a) and \ref{rgsl}(b) show the dimensionless ${\tilde{a}_{l}(B)}/{\bar{a}_{l}}$ versus magnetic field for the $^{85}$Rb$|2, -2\rangle$+$^{87}$Rb$|1, -1\rangle$ and $^{85}$Rb$|2, +2\rangle$+$^{87}$Rb$|1, +1\rangle$ open channels, respectively, for $l = 0,1,2,3$ (note only $l=l^{\prime}$ are considered within our model). All together 5 $d$-wave FRs (3 ``broad" and 2 ``narrow") are predicted in the two channels from 0 to 1000\,G \cite{supplementalmaterial}. Besides the ``broad'' $d$-wave FR ($\zeta_\mathrm{res}=-3.5$) discussed in Fig.\,\ref{splitting}, another ``broad'' $d$-wave FR ($\zeta_\mathrm{res}=-6.7$) is observed in the $^{85}$Rb$|2, +2\rangle$+$^{87}$Rb$|1, +1\rangle$ channel at 622.6(2)\,G \cite{supplementalmaterial}. The latter is of particular interest because it is in the lowest-energy open channel and thus free of two-body dipolar spin-flip loss. Furthermore, two ``narrow'' $d$-wave FRs, one at 337.2(2)\,G ($\zeta_\mathrm{res}=-0.13$) and another at 669.0(2)\,G ($\zeta_\mathrm{res}=-0.16$) [corresponding to blue empty circles in Fig.~\ref{rgsl}(a) and \ref{rgsl}(b), respectively], are found but with a singlet structure for reasons to be discussed.

Despite extensive efforts, we are unable to find any of the predicted ($l=3$) ``broad'' $f$-wave quartets (Fig.\,\ref{rgsl}). This is understood to be caused by the long tunneling time $\tau_l$ required by the entrance wave function to build up within the centrifugal barrier. $\tau_l$ is closely related to the width of a shape resonance, $\gamma_l$, by $\tau_l=1/\gamma_l$. If the length scales associated with the shorter range interactions are well separated from $\beta_6$, and $l$ is not too large, $\tau_l$ follows a universal behavior characterized by $\tau_l = (1/\gamma^{(6)}_{sl})s_t$, where $s_t=\hbar/s_E=2\mu\beta_6^2/\hbar$ is the characteristic vdW time scale. The universal width function $\gamma^{(6)}_{sl}$ is given by (Eq.\,(41) of ~\cite{Gao2009singlechannel})
\begin{equation}
\gamma^{(6)}_{sl} \approx \frac{2(2l+3)(2l-1)\bar{a}_{sl}(\epsilon_{sl})^{l+1/2}}
	{1+2w_l\epsilon_{sl}} \;,
\label{eq:wsl2}
\end{equation}
for a "broad" resonance located at $\epsilon_l$ above the threshold. Here $\epsilon_{sl}=\epsilon_l/s_E$ is the scaled energy, $\bar{a}_{sl}$ and $w_l$ are two $l$-dependent constants~\cite{aslformula,wlformula}.

\begin{figure}
\includegraphics[width=\columnwidth]{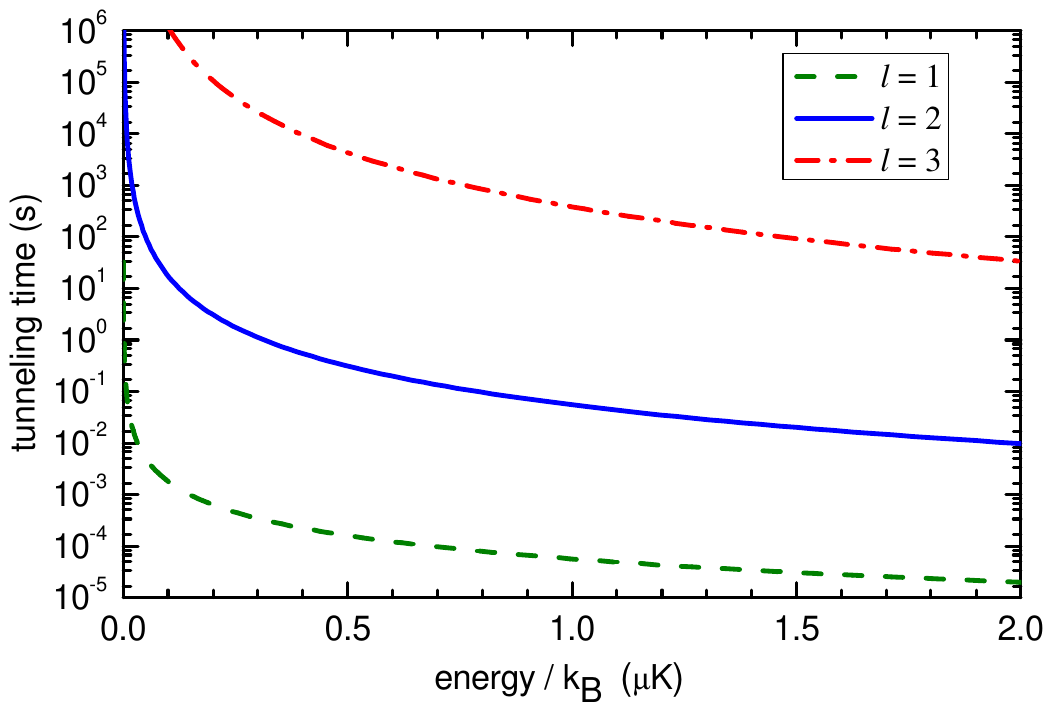}
\caption{(Color online) Tunneling time through the angular momentum barrier, $\tau_l$, for $^{85}$Rb-$^{87}$Rb system, as a function of energy $\epsilon$ in units of $\epsilon/k_B$.
\label{fig:TTime}}
\end{figure}

Figure~\ref{fig:TTime} illustrates, for $^{85}$Rb-$^{87}$Rb system (for which $s_t\approx 1.029\times 10^{-7}$ s), the tunneling time as a function of energy in units of $\epsilon/k_B$, for partial waves $p$, $d$, and $f$. At 400\,nK, the tunneling time for $d$-wave is 0.54\,s which is shorter than the 1.6-s holding time adopted in the measurement for Fig.~\ref{splitting}(d). For temperatures substantially lower than 400\,nK, the $d$-wave resonant coupling do not have sufficient time to establish and the triplet structure cannot be observed. For the $f$ wave, even at 1.6\,$\mu$K, the tunneling time is already 74\,s, much longer than the lifetime of our sample. While higher temperatures shorten the tunneling time, they also broaden the width of the resonance. Therefore we do not expect to observe resolved quartet structure.

The above estimate (Eq.\,\ref{eq:wsl2}) applies only for a ``broad'' FR. For the ``narrow'' $d$-wave FRs at 337.2\,G and 669.0\,G, no triplet structure is observed even at 400\,nK. Our coupled-channel calculations, on the other hand, predict for both of these resonances a triplet splitting resolvable by our setup, but a much smaller width than those of the observed ``broad'' $d$-wave FRs. The observed singlet peak loss feature comes instead from resonance between the $s$-wave open channel and the $m'_l=0$ $d$-wave closed channel. It is actually a $s$-wave FR (scattering) which does not disappear at lower temperatures.

In conclusion, we apply an analytic MQDT to predict and describe FRs in an ultracold $^{85}$Rb-$^{87}$Rb mixture. Two ``broad" $d$-wave FRs are identified, one of them being in the lowest-energy open channel free from two-body dipolar spin flip. By placing the atoms in optical lattice to reduce three-body recombination rate, the latter FR could find important applications in simulating $d$-wave coupling dominated many-body systems.

This work is supported by NSFC (No. 91421305, No.
91636213, No. 11654001, No. 11374176, and No. 11574177), and by the
National Basic Research Program of China (973 program) (No. 2013CB922004
and No. 2014CB921403). Bo Gao is supported NSF under grant Nos. PHY-1306407 and PHY-1607256. We thank Jinglun Li for inputs using coupled-channel calculations.

%\bibliography{reference}


\end{document}